# Refugees and Host State Security: An Empirical Investigation of Rohingya Refuge in Bangladesh

Sarwar J. Minar, Northern Illinois University

**Abstract**

While it is conventionally believed that large scale refugees pose security threats to the host community or state. So, since the massive influx of Rohingyas in Bangladesh in 2017, which resulted a staggering total of 1.6 million Rohingyas refuge in Bangladesh, it was argued that Bangladesh will face severe security threats. This article investigates the security experience of Bangladesh in case of Rohingya influx in a span of three years, August 2017 to August 2020. The research question I intend to address is, has Bangladesh faced security threat due to massive Rohingya influx? If so in what ways? I test four security threat areas, which include, societal security, economic security, internal security, and public security. I have used newspaper reports or newspaper content analysis over past three years along with interview data collected from interviewing local people in cox's bazar area in the first half of 2019 where the Rohingya camps are located. The identity of the interviewees is kept anonymous as per request. In order to assess if the threats are low level, medium level, or high level, I look into both the frequency of reports and the way they are interpreted. I find that Bangladesh has not experience any serious security threat in the last three years. There are some criminal activities and offenses, but these are only low-level security threat at best. My research presents empirical evidence that challenges conventional assertions that refugees are security threats or challenges in the host states.

## I. Introduction

In August 2017, over 725,000 more Rohingyas were forced out of Rakhine State (Myanmar) into Bangladesh (UNHCR 2019). This happened on top of past track record of approximately 200,000 in 1978 and roughly about 250,000 Rohingyas in 1991 fled to Bangladesh from Myanmar (Parnini 2013; Minar and Halim 2020). The results a total of 1.6 million of Rohingyas living in Bangladesh (S. Alam 2019). The Rohingyas are considered as the single biggest stateless community in the world (Advisory Commission on Rakhine State 2017). The Rohingya developed through a long historical complicated process in the Rakhine State (Minar and Halim 2020) but they have not been recognized by the military led government of Myanmar. Even the term Rohingya did not get current until before the independence of Myanmar (Minar and Halim 2019). The Tatmadaw's latest episode of brutal crackdown on the Rohingyas can be understood both from internal and external reasons (Minar 2019). However, while it is conventionally believed that large scale refugees pose security threats to the host community or state. Therefore, scholars argued that with the large scale Rohingyas influx into Bangladesh, the country was going to face wide range of severe security threats. This article tests the claim by empirically investigating the security experience of Bangladesh due to Rohingya protracted refuge over a span of three years, August 2017 to August 2020.

The research question I intend to address is, has Bangladesh faced security threat due to massive Rohingya influx from 2017 to 2020? If so, in what ways? I test four security threat areas, which include, societal security, economic security, internal security, and public security. I assess the four security threats in three categories: low level, medium level, and high-level threats. I investigate both the frequency of reports, nature of event/issue occurred, and the way they are interpreted. I find that Bangladesh has not experience any serious security threat in the last three years. There are some criminal activities and offenses, but these are only low-level security threat at best.

This article is divided into seven parts. Following introduction, second part presents an overview of the emerging popular discourse of Rohingya induces security threat in Bangladesh. The third part presents the research methodology, the fourth part lays out the theoretical framework. In the fifth part I present my investigation of four security threat areas in accordance with the theoretical framework. The sixth part provides a quick overview of the findings. Seventh part draws some limitations of the study and then I draw a conclusion.

## II. Rohingya refugees induced security threat in Bangladesh: an emerging popular discourse

With the latest inflow of the Rohingyas in Bangladesh a popular discourse emerged in Bangladesh that the Rohingyas stay is likely to power [national] security threats to Bangladesh. This prognostic security threat discourse emerged through the commentaries of the top government officials, many civil society stake holders, and decent size of scholarly work. Many government officials frequently publicly claim that Rohingyas pose national security threats to Bangladesh. The Foreign minister for instance say that the Rohingyas may pose national security threat to the county (Desk 2017; Report 2017). Gowher Rizvi, the international affairs adviser to the Prime Minister of Bangladesh says, Rohingya crisis will become biggest security threat to South Asia (Tribune Desk 2019a). The Prime Minister herself voice similar concern that the Rohingyas to be national and regional threats (UNB 2019).

Some scholars also come up with research that finds Rohingyas to pose national security threat to Bangladesh in multiple levels. Utpala Rahman (2010) argues that the Rohingyas present political, social, economic, environmental *security dilemmas* for Bangladesh as it presents dilemma between its national interest and human security of Rohingyas (U. Rahman 2010). The author further fins that Rohingyas to be potential threat to Bangladesh's internal stability (Rahman 2010). Mayesha Alam (2018) illustrates political, legal, economic, social, and environmental effects of the Rohingya influx in Bangladesh (Mayesha Alam 2018). Lailufar Yasmin and Sayeda Akther (2019) found Rohingyas to be threat to Bangladesh on grounds like, demography, social tension, law and order, (forced) marriage, prostitution (HIV etc.), environmental (Yasmin and Akther 2020). Hossain Ahmed Taufiq (2019) find Rohingya to cause insecurity in Bangladesh on grounds of social, economic, environmental, health sectors (Taufiq 2019b). In this connection, Mills and Norton (2002), for instance, in assessing refugees' security in great lake region of Africa, considers four security areas: human security, societal security, state security and international security (Mills and Norton 2002).

## III. Methodology

I test four security threat areas, which include, societal security, economic security, internal security, and public security. I have used newspaper reports or newspaper content analysis over past three years (August 2017 to August 2020) along with interview data collected from interviewing local people in cox's bazar area in the first half of 2019 where the Rohingya camps are located. If there is no news report or none of the interviewees talk about certain security area, I consider there is no security threat. In order to be included in the analysis, a security threat into analysis at least one interviewee has to talk about it or at least one conflict incident or once such incident has been reported to the news media or at least one interviewee mentioned that as threat as particular type of security.

I analyzed news contents of two top daily newspapers in Bangladesh (widely known and highest circulation): The Daily Star and The Dhaka Tribune. I searched their online archives to get access to newspapers copies form August 17, 2017, to August 17, 2020.  I focused on the news stories and/or covering containing 'Rohingya' and/or 'Rohingya as security threat' within the newspapers archival record. I checked all the *titles* of the news published in the newspapers and recorded all *news content* having any type of wording that reflects security concern or security threat for Bangladesh, be it commentary, opinion pieces, editorial, and any report regarding incidence of violence. In total 57 news reports were found.

Additionally, I collected police complaint records or filed cases from the local police station that has Rohingyas involvement or crimes committed by Rohingyas. The collection of police cases remains limited since due to a recent initiative of reorganization of whole police force deployed in the cox's bazar area created a very unfavorable situation to get more such records.

In order to assess and determine if the threats are low level, medium level, and high level, I looked into both the frequency of reports/news with similar category of incidents as well as well the way the incidents/news are interpreted/described. I explore details about conflict incidents/ incident of violence including the nature of conflict on news reports, whether conflict incidents are just got reported on the news media, or the reports the conflicts incidents are interpreted as low-level, medium level or high-level security threat (See Appendix B).

By *low level security threat*, I refer to incidents that affects only an individual or small groups of up to 15 people, incidents that have very limited adverse effect which has only local implication, the incidents signify threat type which is usual or same as happen in the other parts of the country, for example, petty crimes, person to person tussle etc., and is easily solvable by local authorities. By *medium level security threat*, I refer to incidents that affect small group of 15-30 people (Group level incident), has limited adverse effect (only local implication), the threat type is usual same as other parts of the country is facing, for example, petty crimes, person to person tussle etc. (the scope may be bigger), and is solvable by local authorities and law enforcement agencies. And lastly, by *high level security threat*, I refer incidents that affect large groups of more than 30 people, has adverse effect that goes beyond local context and have national level implication, is unprecedented/unique type of threats with possible national level implication for law-and-order situation, and is challenging for the law enforcement agencies to solve it (See Appendix B).

## IV. Theoretical Framework

I do not intend to discuss the theoretical and conceptual development of security (e.g., traditional security and non-traditional security etc.) and which one is more important and relevant than the other etc. because that is not the focus of this article. So, I discuss only the security issues that I consider for investigation in this paper. However, shortly, the traditional security focuses on the 'states' as referent object of analysis and therefore state security, military security etc. became dominant areas of research. Whereas with the emergence of non-traditional security in the post-Cold War era, now 'individuals' are considered the referent object of analysis (Hama 2017) and therefore all issues that poses threat of humans have become focus of research attention. The latter is more appropriate to investigate security threats for host community emanating from refugees. Relatedly, I consider refugees flow as the case of large-scale forced migration (LSFM), which is a special case of population displacement that affects a sizable portion of a country's population at least 5 percent (Bayar and Aral 2019).

I consider four security areas in this paper: Societal Security, Economic Security, Internal Security, and Public Security (See Appendix A). Julia Tallmeister uses these four security threat areas to investigate whether immigration is a threat to security for European states (Tallmeister 2013). I apply these to investigate refugees induced security threat to the host community. The four security areas more or less fall within the category of non-traditional security research agenda. In assessing host state security due to refugees' flow, various authors used various frameworks. Mills and Norton (Mills and Norton 2002), for instance, in assessing refugees security in great lake region of Africa, considers four security areas: human security, societal security, state security and international security. Using human security is confusing as it seems to imply human security of both the refugees and host community. Using levels of analysis (e.g., individual, national, international) is another way of investigation but again the individual category sems to integrate both refugees and host community people. Maisha Alam (Mayesha Alam 2018) uses political, legal, economic, social, and environmental effects of the Rohingya influx in Bangladesh. Some of her points are excellent areas but others (e.g., political, legal) are not compatible with host state security assessment. Therefore, I consider societal, economic, internal, and public security to be more relevant to understand host state, in this case Bangladesh's, (in) security scenario due to massive refugees or Rohingyas influx.

By societal security, I simply refer to the collective national identity of host state, in other words, cultural, linguistic and identitive survival of particular social group (Theiler 2003). Societal security is defined as the defense of an identity against a perceived threat, or more precisely, the defense of a community against a perceived threat to its identity (Wæever 2008). In this case host community. From this perspective, the national values are considered to be under threat (Weiner 1992). I investigate whether Rohingyas refuge pose any challenge to the national identity of the host state, Bangladesh, as they have foreign culture, language, or religion etc. (See appendix A).

By economic security, I refer to the ability of individuals, households or communities to cover their essential needs sustainably and with dignity. The essential needs include, food, basic shelter, clothing, health care, and education (related expenditure). The essential assets needed to earn a living, and the costs associated with health care and education also qualify (ICRC 2015). For ILO economic security is composed of basic needs infrastructure pertaining to health, education, dwelling, information, and social protection, as well as work-related security (ILO n.d.) which complements the definition. From this perspective, the economic capability of host community people is considered to be under threat. I investigate whether Rohingyas refuge pose any challenge to the host community people ability to meet essential needs or employment on which the ability depends on (See appendix A)..

By internal security, I mean the process of maintaining peace within the boundaries of a sovereign state and other self-governing territories, usually done by maintaining national law (Karthikeyan 2019). From this perspective, country's internal law and order situation is considered to be under threat. I look specifically into disruption of law-and-order situation, terrorist/extremist/insurgency activities (e.g., recruitment, communication, training etc.), smuggling, illicit drug business, small arms business, criminal offenses causing security and order disruption in the locality or in the country etc. (See appendix A).

And, by public security, I mean the protection of the public or safeguarding people from crimes, disaster, and other potential dangers and threats (Goodwin University 2019). From this perspective, the wellbeing of the people is considered to be under threat. Though wellbeing is a fluid and contested concept, I simply refer to the state of being comfortable, healthy, and happy. I specifically investigate issues of criminal activities affecting people's wellbeing, smooth or disruption to communication facilities (e.g., road transport facilities) and similar threats and dangers affecting peoples' life. I integrate health security within public security instead of creating a separate category. So, therefore, illicit drugs usage in local community/country and its impact, other health risks are within this category (See appendix A)..

## V. Assessing Security Threat Experience of Host Community: The Case of Rohingyas in Bangladesh

How do refugees threaten host state's national security? With different socio-cultural belief, the refugees threaten national socio-cultural cohesion (collective identity) of the host state. They also threaten the host state economy by disrupting host state labor market and the economic capacity of the host state people by adding extra stress to the job and labor market. Additionally, as they refugees get involve with various crimes and law-and-order disruptive activities, they threaten internal security of the host state and with extra pressure cause imbalance of social and public order (Wang 2012) . Loescher argues that that Refugee movements not only create and exacerbate conflict between states, but influence their relations in many other ways as well (Loescher 1992a, 4). Weiner points that how and why some migrant communities are perceived as cultural threats is a complicated issue, involving initially how the host community defines itself (Weiner 1992). It is pretty common impression created by the news that Rohingyas created potential for instability in the localities as well as other parts of the country has led to labor insecurity, food inflation, water shortage, environmental damage and also anti-social activities (Tribune Desk 2020). President of Civil Societies Forum Cox's Bazaar Fazlul Kader, for example said, "In the beginning every one of our localities welcomed the displaced Rohingyas on humanitarian ground.

We worked for their food, clothes, and shelter. But, the situation has changed now as they have already become a burden for us" (Tribune Desk 2020).

This section maps the four security areas and experience of Bangladesh due to Rohingyas from the newspaper reports. While the general impression created by the news is that the Rohingyas placed increased burden locally, nationally, and internationally from areas which include economic, political, societal, internal, and public security, this section fleshes out the security experience and related findings.

## A. Societal (In) Security:

As previously mentioned that in the post-Cold War era, under human security framework individuals are considered as sole referent object of security (Hama 2017). Individuals live in society. Society is defined as the social unit that provides the primary locus of identification for its members. Objectively, a society is signified and differentiated from other societies by markers such as language and customs; and subjectively, it is the repository of shared meanings and identifications for its members who possess a 'we-feeling' (original emphasis by Karl Deutsch) (Theiler 2003). Both Weiner and Loescher in their respective research depicted that refugees can pose threat to the host state national security in numerous ways and threat to socio-cultural unity of host state is one of the worst one (Weiner 1992; Loescher 1992b). Weiner's research showed that refugees (and migrants) are considered as threat to host states' cultural identity (Weiner 1992, 110). Loescher's research reveals that refugees (and immigrants) pose threat to social complexion of the host country (Loescher 1992b, 48). In this study, I specifically investigate Rohingyas merging with Bangladeshi people, for instance, spreading through local community, marriages (though banned), spreading throughout the country, gaining legal status through illegal means (e.g., acquiring National Identity Cards, Passports etc.) as indicators for socio-cultural threat to Bangladesh (See Appendix A).

Bangladesh is a homogenous state comprising of similar ethno-linguistic people, Bengalese. Though there are some indigenous people living in the country which include Chakma, Marma, Garo, Shantel and others, the collective national identity is Bangladeshi (i.e., the citizens have a 'we-feeling'). With a very little divergences, the majority of the people hold same cultural and linguistic identity. All the people hold the same national identity and national value. With the Rohingyas influx and protracted refuge in Bangladesh, many consider Rohingyas as societal threat. Rohingyas have different culture and language (i.e., 'others'). The concern is that as they outnumbered the locals and are gradually not only getting mixed with the locals in cox's bazar but also spreading throughout the country and therefore posing a threat to the national value and national identity of Bangladesh.

Though the Rohingyas are not allowed to leave the refugees camps, they often go out of the camp. The old Rohingyas (who came to Bangladesh before 2012) have developed 'know how' about how to get out of the camps irrespective of time.[1] Moreover there are many unregistered

---

[1] Interview, Local dweller, Kutupalong Camp area, Cox's Bazar, March 2019

Rohingyas who move free around. Locals belief that due to presence of the Rohingyas the societal fabric is at risk in the Cox's bazar, the area where Rohingya camps are located.[2]

First, the local people in cox's bazar considers Rohingyas as a threat to their communal and social identity.[3] Many locals of Cox's Bazar's Teknaf and Ukhiya Upazilla are feeling unsafe because of the presence of thousands of Rohingyas and they link an increased Rohingya presence to social degradation and criminal activities in their areas (Shaon 2018). They think that due to the inability of the Rohingya to find legal employment to cover their needs, they are turning to illicit trade and acts like robbery, drug trafficking, and prostitution (Shaon 2018). A 30-year-old female resident of Teknaf's Nhilla union said: "Most Rohingyas are illiterate—hence some notorious local people encourage them to get involved in crime. As a result, crime has increased in our society." A 27-year-old local of Teknaf said the recent Rohingya influx has deteriorated culture and morals. (Shaon 2018). These also hint the gradually involvement of the Rohingyas with local affairs.

Second, there has been reports of Rohingyas to spread across the county. For instance, twelve Rohingya refugees were found in Tahirpur Upazilla of Sunamganj, around 600 kilometers from Cox's Bazar camps. Alarmingly, six of them had original Bangladeshi citizenship certificates (Star Online Report 2017a). Another report mentions that the Police arrested 43 Rohingyas, including children, at separate places in Comilla which is about 400 kilometers away from Rohingyas camps (Masud Alam 2019).

Third, another cause of concern is marriage. Though the marriage ban was issued by the Ministry of Law, Justice and Parliamentary Affairs in July 2014 to prevent Rohingyas from assuming Bangladesh citizenship by marrying, there are instances of marriage. Shoaib case is one such example. A Bangladeshi boy fell in love with a Rohingya girl Rafiza and in defiance of the ban on marriage, got married (M. Rahman 2017). Another Rohingya youth named Ziabul hacked a local youth named Abdul Jabbar to death at Ramu Upazilla of Cox's Bazar over an alleged extra-marital affairs (Mahmud 2017c) which implies a similar kind of trend in development.

Fourth, there has been evidence that Rohingyas are getting National Identity Cards and Passports through illegal means. One such case is Mosarraf (husband) and Khadiza (wife). Khadiza fled to Bangladesh in 2002 and two of their sons have Bangladeshi NID cards. Khadiza's younger brother Shahidullah, 23, has been studying at a Cox's Bazar madrasa since 2013. He said many from their village had come to Bangladesh for study and got documents. Another Rohingya man, Selim, who fled from Buthiadaung (Myanmar) area, revealed that he had travelled to Malaysia on a Bangladeshi passport. However, with the introduction of Machine-Readable Passports, it has become difficult to get passport illegally (Mahmud 2018b).

If we analyze the news content report findings, I find twelve news reports from August 2017 to August 2020 that involve societal security threats. There are five reports in first year, seven in second year and one in the third year. The incidents reported are mostly petty crime and personal level tussle as seen in other part so the country not having refugees, and all have impact on less than 15 people thresholds, they have only local level implications, and these are solvable by local law enforcement authorities. So, these incidents fail to reach the medium level security threshold let alone high-level security threats. That is why I conclude that these incidents reflecting societal security threats can best be described as a low-level threat both due to less frequent reports and the way this has been explained till now.

---

[2] Interview, Local Political Leader, Kutupalong Camp area, Cox's Bazar, March 2019
[3] Interview, Local Businessman, Kutupalong Camp area, Cox's Bazar, March 2019

## B. Economic (In) Security:

As I mentioned above, by economic security I mean the ability of individuals, households or communities to cover their essential needs (e.g., food, basic shelter, clothing, health care, and education etc.) sustainably and with dignity (ICRC 2015). Both Weiner and Loescher in their research illustrated that refugees can pose economic threat to the host state (Weiner 1992; Loescher 1992b). Weiner's research shows that refugees (and immigrants) create a substantial economic burden by straining housing, education, and transportation facilities (Weiner 1992, 114). Weiner further shows that refugees may illegally occupy private or government lands; their goats, sheep, and cattle may decimate forests and grazing land; they may use firewood, consume water, produce waste etc. (Weiner 1992, 114). Loescher argues with refugees (and immigrants) the economic tension rises as the influx of labor drive down wages and create unemployment while driving up the cost of housing and other goods (Loescher 1992b, 48). In this research, I specifically investigate the financial ability or the economic capability of Bangladeshi people (e.g., job, business, price hike etc.) to meet their needs and how those abilities are affected due to Rohingya refugee situation. I analyze the economic security situation from three angles: local, national, and international.

First, from local perspective, the local peoples, those who live in the areas where the Rohingya refugees' camps are located, are most adversely affected in a number of ways: loss of crops and cultivable land, loss of jobs, decrease of labor cost/wages, price hike of necessary goods and transport etc. The local people, especially those who are at the bottom, are worst affected. Some of them lost their crop and cultivation land, and some other lost their jobs. On top of that their life is negatively affected due to increased cost of necessary goods and other expenses than usual. Kabir Ahmed, a local who has been living in Ukhiya's Baluchhara for the last two decades said that the Rohingyas occupied huge arable lands, he said. "We had five acres of land but the Rohingyas have grabbed everything" (Aziz 2018). Madhurchara village's farmer Baset Ali said the Rohingya refugees had destroyed his paddy fields, in his words, "I was compelled to sell my cows to feed my family. The Rohingya have also destroyed my trees planted under social forestation project." Another local dweller, Rashida Begum, said, "Our crops were destroyed after the Rohingya came. We are helpless now and have been forced to beg" (Mahmud 2017b). Mohammad Kobir also expressed that he was worried about his livelihood after fishing was temporarily banned in the Naf River after the influx. The Rohingya influx totally disrupted the daily activities of the residents, harvesting, fishing activities, border trade, service, and daily commerce could come to a standstill (Shibli 2017). A local businessman objected that he was facing unequal competition in his business since a Rohingya started the same business in his locality without government permission.[4]

Local people were losing jobs or getting paid less than before, on the other hand the prices of essential commodities soared due to high demand, local produces were not in demand as goods from outside flooded the camps, and people have lost their livelihood from cultivable land that had to be surrendered for the refugee camps (Tribune Desk 2020). Local suffered due to food and transport price hikes, food grain shortages, and reduced tourism (Husain and Ovi 2017). Many locals had job related to tourism industry (e.g., photography, bike, and horse riding etc.). Rohingya refugees spreading throughout the district threaten tourism as tourists were reluctant to visit over security issues and fears of possible chaos (Husain and Ovi 2017). Many local residents, near makeshift refugee camps in Ukhiya Upazilla in Cox's Bazar, have also expressed their concerns over the climbing prices of essential commodities. "The prices of almost everything has risen in the last two weeks due to higher demand following the Rohingya influx," said Shamsul Alam, a resident of Balukhali. He further said that onions were selling at Tk60 per kilogram and potato at Tk40 per kg, compared to Tk45 and Tk25 respectively in Dhaka, where goods' prices are always higher than local areas  (Husain and Ovi 2017). Prices of some daily commodities had more than doubled at several local markets in Ukhiya. Rafiqul Islam, a trader, claimed that a syndicate of unscrupulous businessmen had hiked the prices soon after the influx. Commuting has also become costlier. Two months ago, it cost about Tk200 to hire an autorickshaw to travel to Teknaf from Cox's Bazar but now (immediately after 2017 influx) it cost nearly Tk 500. The Cox's Bazar-Ukhiya-Teknaf road is now crammed during peak hours due to relief transport, movement of VIPs and administrative vehicles (Mahmud 2017b) and the Bus fare has also increased (Mahmud 2017b) which put extra pressure on financial ability of the local people.

Second, from national perspective, there has been a big cost.  Around 6,000 acres of land has already been deforested in Ukhiya and Teknaf by the Rohingya camps and the estimated value of the land is Tk 741.31 crore ($86.67 million) (Star Business Desk 2018). "The Rohingyas have occupied all forest lands and hills in the area," said Kabir Ahmed, who has been living in Ukhiya's Baluchhara for the last two decades (Aziz 2018). The Rohingyas have occupied a 10-km-long stretch of forest near Kutupalong market. The Rohingyas have occupied 4,000 acres of land though the government has designated 2,000 acres for them. Local youth, Helaluddin, said the Rohingyas not only occupied their homesteads but also cut down all the trees and hills. Md Ali Kabir, divisional forest officer of Cox's Bazar (south) forest department, said the Rohingyas have grabbed 10,000 acres of land in Ukhiya range.  The old Rohingyas (who came before 2012) occupied 6,000 acres of land and the new ones claimed 4,000 more acres of land  (Aziz 2018). An estimate sows that the  600,000 newly arrived Rohingyas settled in Teknaf and Ukhia upazilas, are costing more than USD 4 million daily to feed and shelter, not counting the other half a million who came earlier (Shibli 2017). However, besides the negative aspects, there is an opposite side if we look at national scenario. The crisis also created job opportunities for many Bangladeshis in NGOs sectors working on the issue. Some industries got economic

---

[4] Interview, Local Businessman, Kutupalong Camp area, Cox's Bazar, March 2019

benefit as well. The hospitality and airline industry, have witnessed a surge in demand from journalists, international observers, and human rights activists who are flocking to the Cox's Bazar region to report on the crisis, observe firsthand the condition of the refugees, and to offer assistance to the needy (Shibli 2017).

Third, from international perspective, Bangladesh must worry about when it comes to economic dimension. There is a big labor market for Bangladesh in the Middle East from which Bangladesh receive handsome remittance. In a study in 2010 Utpala Rahman found that (U. Rahman 2010) as many Rohingyas were going to middle east taking Bangladeshi passports through unfair means and they were involved in activities tarnishing the image of Bangladesh that was a threat to Bangladesh's economy. My study finds that the trend continues in 2020. Although the government holds no data on the exact number of false passports in circulation, Expatriates Welfare Minister Nurul Islam said in April 2018 that about 250,000 Rohingyas had gone abroad with Bangladeshi passports. The minister said Bangladeshi workers are facing an "image crisis" as a result. "Many of them are involved in criminal activities abroad" (Mahmud 2018b). Even 13 Rohingya men with Bangladeshi passports were deported to Dhaka from Kingdom of Saudi Arabia. Former Expatriates Welfare Minister Nurul Islam said in April 2018 that about 250,000 Rohingyas had gone abroad with Bangladeshi passports (Rabbi 2019).

If we analyze the news content report findings, I find six news reports from August 2017 to August 2020 that involve economic security threats. There are two reports in first year, three in the second year and one in the third year. The incidents reported are mostly petty crime and personal level tussle as seen in other part so the country not having refugees, and all have impact on less than 15 people thresholds (single incident), they have only local level implications, and these are solvable by local authorities. The national scenario reflects an opposite scenario, it has created new job opportunities for the Bangladeshi nationals. So, on economic grounds, the incidents fail to reach the medium level economic security threshold let alone reaching the high-level security threat. That is why I conclude that these incidents reflecting economic security threats can best be described as a low-level threat both due to less frequent reports and the way this has been explained till now.

### C. Internal (In) Security:

Generally, internal security refers to the threats to law and order in the country from sources within the country (Grizold 1994). However, as I mentioned before, by internal security, I mean the process of maintaining peace within the border of a sovereign state and other self-governing territories, usually done by maintaining national law (Karthikeyan 2019). Jacobsen's research showed the trace of gun running, drug smuggling, and human trafficking due to refugees stay on host country (Jacobsen 2002). In this paper, I specifically investigate how Bangladesh's internal law and order situation, internal peace and tranquility, peaceful coexistence of communal harmony etc. (Karthikeyan 2019) are affected or disrupted due to the Rohingyas refugees' protracted stay in the country. I tried to look into if there is any news about terrorist/extremist/insurgency activities (e.g., recruitment, communication, training etc.), traces of organized crimes, smuggling, illicit drug business, small arms business, criminal offenses causing security and order concerns causing disruption in the locality or in the country or threaten the functioning of country or parts of the country.

First, there have been reports saying the unemployed and idle Rohingyas are being drawn into criminal activities for money thought they don't have to think about their food and accommodation as the aid agencies are providing them with everything they need. So, a big portion of Rohingya youths remain idle and become engaged in domestic violence, internal feuds, and gender-based violence (Mahmud 2018a).

Second, there has been news regarding Rohingyas involvement with clash with law enforcement agency members and local people. A Police got beaten up by a Rohingya couple beat up a policeman at Hnila while a local resident was killed by a refugee (Mahmud 2017b). In Balukhali Camp, a gang of Rohingya men attacked and beat up a group of five local laborers and accused them of robbery. Law enforcers detained two Rohingya men with two guns and six bullets in connection with the incident (Mahmud 2017c). There are also reports that the Rohingya have attacked a relief volunteer in Ukhiya (Mahmud 2017c) and foreign journalists in the camps at Ukhiya and Teknaf (Aziz 2019).

Third, a number of extremist and militant groups tried to take advantage of the Rohingya crisis have taken to social media in order to urge their followers to actively participate in violence in the Rakhine state of Myanmar in support of Rohingya refugees. All of the pages also posted a recent two page statement by the General Leadership of Al-Qaeda regarding the formation of a fund to aid Rohingya refugees, which further called for militant groups to train and set out for the Rakhine state (Mahmud 2017a). Another news reveals that a top Counter Terrorism and Transnational Crime (CTTC) official said the JMB (a terrorist organization in Bangladesh) men were working in the Rohingya refugee camps since 2016 and were recruiting members behind the relief and funding aid to the Rohingya camp. They recruited at least 40 militant members from there (Rabbi 2018). The CTTC official also said that members of the militant organization believe that elections and democratic methods are "Kufri Motobad" which means disbelief doctrine. Planning of targeted killing of political local level leaders (Rabbi 2018). Another report posits that lack of formal education in the Rohingya refugee camps left youths susceptible to radicalization and a Chattogram-based Islamist movement Hefazat-e-Islam, which publicly called for jihad against Myanmar, has considerable influence over the madrasa network in the camps, through the funding and religious scholars that it provides (Diplomatic Correspondent 2019). However, there was no follow up of similar news. Deputy Commissioner (Cyber Crime) Mohammad Alimuzzaman ensured that they were monitoring the situation and trying to trace the IDs of the people who run the pages on social media. Bangladesh Home Minister Asaduzzaman Khan Kamal also ensured that Bangladeshi security agencies proved their capacity and were in high alert and prepared to ruin any plans of militants anywhere in the country (Mahmud 2017a).

Fourth, Rohingyas have been found to be involved with illicit drug business. New reports covered that many Rohingyas are smuggling yaba pills from Myanmar's border areas into the camps for storage and then they send on to Cox's Bazar and other parts of Bangladesh (Mahmud 2018a). Rohingya yaba carrier Motaleb Mia (pseudonym) said he knows 20-22 Rohingyas who are involved in yaba smuggling and explained that "The smugglers hand yaba pills to the mules in the deeper parts of the camps, who carry them to Cox's Bazar". A report in October 2017 reveals that law enforcers, in different drives, have seized yaba tablets worth around Tk13 crore in the two upazilas, and have arrested several Rohingya for suspected involvement with smuggling (Mahmud 2017c). Another report revealed that Security forces seized more than 10 million yaba pills from Rohingyas and local drug peddlers since the beginning of the latest refugee influx till October 2018 (Mahmud 2018a). interestingly, another report reveals that some Rohingyas are involved in drug-peddling work for NGOs or independent workers (Mahmud 2018a). Another report mentions that 68 narcotics-related cases were filed by 2018 (Chandan and Rubel 2018).

Fifth, news reports show that Rohingyas have been found to possess small arms and in small arms business. One report says that two Rohingya men, Nurul Boshor and Md Elias, were caught by law enforcement agency and found firearms in their possession (Mahmud 2017c). Police said many Rohingya youths are involved in arms dealing and robberies. So far, 12 cases have been filed and several arrests have been made. The Rohingyas, in collaboration with local dealers, get involved with the arms trade in Maheshkhali - the hub of Bangladesh's local firearms manufacturing. The police's counter-terrorism units have information that there are five to six Ak47 rifles circulating in the Rohingya camps. Robbery occurs in villages surrounding the Nayaparha and Leda refugee camps in Teknaf. Rohingya robber gangs are frequently engaged in infighting. At night inside the camps, gunshots are often heard (Mahmud 2018a). Dr Gowher Rizvi PM's foreign policy adviser in a seminar mentions that weapons seem to come to this region which may cause instability in the region (Tribune Desk 2019a).

Sixth, there has been an increasing rate of Rohingyas involvement in criminal activities. During this period, police arrested more than 1,000 Rohingyas for various crimes which also include smuggling, human trafficking, and prostitution. A total of 563 Rohingyas have been sentenced (Mahmud 2018a). Police arrested about 300 Rohingya in cases involving killings, robberies and abductions in the camps since by October 2018 (Reuters 2018). By the end of first year (August 25, 2018), 485 Rohingyas were accused in 250 cases along with 80 drug cases (The Daily Star 2018). By the end of second year (26 August 2019), about 328 cases have been filed against 711 Rohingyas over various crimes (Molla 2019). By the end of third year (August 20, 2020), the number increased to a total of 725 criminal cases involving Rohingyas have been filed against 1,664 individuals (Rashid 2020).

Seventh, the law and order situation is getting worse day by day mostly because of internal issues, clashes between Rohingya groups over past disputes and attempts to establish dominance, both in the camps and over the host community (Aziz 2019). Rohingya camps in Ukhiya and Teknaf have more and more drug smuggling, violent crime such as murder, vandalism, yaba smuggling, engaging in violence and attacking the police and so on (Mahmud 2017c). Arifullah, an appointed a leader of thousands of refugees got stabbed to death by a group of Rohingya men (Reuters 2018). Many murders inside the camp are linked to rivalry and grouping among majhis (Chandan and Rubel 2018). Such violent criminal activities remain limited within camps and have limited effect on host community.

If we analyze the news content report findings, I find fourteen news reports from August 2017 to August 2020 that involve internal security threats. There are four reports in first year, seven in second year and three in the third year. The incidents reported are mostly petty crime and personal level tussle, illegal drug carrying activities, as seen in other part of the country not having refugees, and all have impact on less than 15 people thresholds, they have only local level implications, and these are solvable by local law enforcement authorities. So, these incidents fail to reach the medium level security threshold let alone high-level security threats. So, I conclude that these incidents reflecting internal security threats can best be described as a low-level threat both in terms of frequency and they are interpreted till now.

### D. Public (In) Security:

By public security, I mean the protection of the public or safeguarding people from crimes, disaster, and other potential dangers and threats (Goodwin University 2019). In this study, I investigate how the wellbeing of the people is being affected due to Rohingyas refuge in Bangladesh. I specifically investigate issues of criminal activities affecting people's wellbeing of day-to-day life, for instance, disruption to communication facilities (e.g., road transport facilities), education facility etc. affecting peoples' life. I integrate health security concerns within public security instead of creating a separate category. So, therefore, illicit drugs usage in local community/country and its impact and other health risks (e.g., HIV etc.) are integrated within this category.

First, the local public has become increasingly concerned about high crime rates intensified by the Rohingyas and the threat it poses to public wellbeing and social order. Palangkhali Union Chairman M Gofur Uddin Chowdhury said, "The locals are suffering the most" (Mahmud 2017b). There have been reports of the refugees getting involved in illegal activities, raising security concerns among the locals affecting their life and day to day activities (Mahmud 2017b). News already spread that a local resident was killed by a refugee (Mahmud 2017b) and in Balukhali Camp, a group of Rohingyas attacked and beat up five local laborers (Mahmud 2017c).

Second, the arrival of the refugees has also affected educational activities as many institutions are allowing the local administration and security personnel to use their rooms, said Ukhiya Degree College teacher Shahid Uddin. "The students, too, cannot focus on studies because of the ongoing crisis," he said. Political science student Arman Babu said their classes had been suspended as many students and teachers were involved in relief distribution. (Mahmud 2017b). Also, many local schools and madrasas are used as camps for: law enforcement agencies, medical teams, relief centres, and other purposes. This is why the number of students has declined—which worries the guardians. Kutupalong Government Primary School's Headmaster Habibur Rahman said most of the classrooms are used to provide the Rohingyas with humanitarian services. "A medical team has been set up in a classroom and two classrooms are used as barracks for policemen and Ansar members. Additionally, cooking ingredients are stored at one corner of the school, and a stove has also been set up in the field to cook for Rohingyas (Aziz 2018).

Third, the communication facilities have been negatively affected which in turn caused harm to the wellbeing of the locals. Cox's Bazar-Ukhiya-Teknaf road is now crammed during peak hours due to relief transport, movement of VIPs and administrative vehicles (Mahmud 2017b).

Fourth, there is an increasing risk of HIV infections or spread. October 2017, Health officials in Cox's Bazar fear spread of HIV/AIDS in the refugee camps as a rising number of Rohingyas have been diagnosed with the disease. So far, 19 patients, mostly males, have been identified (Molla and Jinnat 2017). By November 2017 the number of HIV positive Rohingyas rose to 83 Rohingyas, 29 are males, 41 are females, while six children infected with the virus also were found (Star Online Report 2017b). Under the programme, a minister said, various steps have been undertaken to prevent communicable and non-communicable diseases like HIV/AIDS & STD among the Rohingya people staying at different places of Ukhiya and Teknaf Upazillas in Cox's Bazar district (Star Online Report 2017b). A gynaecologist working in the field said that the risk of spreading of the infection is very high, the said, fearing a rapid spread of the contagious disease if steps were not taken immediately (Molla and Jinnat 2017). Hsan and et al. posits that due to absence of basic living facilities during humanitarian crises places girls and women at particularly high risk of exploitation and exposure to HIV and other sexually transmitted infections (Hsan et al. 2019). The prevalence of HIV among injection-drug users in Bangladesh is $1\cdot1\%,2$ and because Cox's Bazar is well known for drug trafficking, both the general and refugee population are at risk of drug abuse and HIV exposure. Taking advantage of the displaced people, the traffickers trick vulnerable ones into sex slavery, domestic servitude, and forced labor. Teenage girls of between 15 and 19 are at a higher risk of being sold into sex slavery. Unprotected sex exposes the customers to the sexually transmitted diseases that these women might be carrying. And spread of these diseases remains a potential threat for the region. According to the US state department's Trafficking in Persons Report 2019, Rohingya girls are transported to Dhaka, Kathmandu and Kolkata, among other destinations, where they are forced into prostitution. Some others become involved with the sex trade closer to home—in Chattogram, and even in the prostitution business in Kutupalong itself. According to a report by *South China Morning Post*, in Kutupalong the "sex industry is thriving". "At least 500 Rohingya prostitutes live in Kutupalong," the same report quoted a fixer as saying while elaborating the sex trade situation in the biggest Rohingya camp in Cox's Bazar (Tayeb 2019).

If we analyze the news content report findings, I find fourteen news reports from August 2017 to August 2020 that involve public security threats. There are five reports in first year, seven in second year and two in the third year. The incidents reported are mostly petty crime and personal level tussle as seen in other part so the country not having refugees, and all have impact on less than 15 people thresholds, they have only local level implications, and these are solvable by local law enforcement authorities. So, these incidents fail to reach the medium level security threshold let alone high-level security threat. The health security risk seems of medium level threat, but then the reports then dropped. That is why I conclude overall public security remains as low-level threat both due to less frequent reports and the way this has been explained till now.

## VI. Key Findings and Discussions

Three years newspaper content analysis reveals that all the four security threats areas, societal, economic, internal, and public, are in low level. The societal security threats remain in the low-level both due to less frequent news reports, 14 reports in 3 years and the way explained in the news. The economic security threat experiences also remain low with 6 reports in 3 years, So, both remain in low level threat as well and again because of the less frequent news coverings and the way interpreted. The reports regarding internal security threat experience shows 14 reports in 3 years, which seem to be a little more dangerous in comparison to first two, however, they still fall under low threat category according to the matrix I used (see appendix B). Lastly, the reports related to public security threats shows 14 reports in 3 years, which is also low level. The health security risk seems of medium level threat, but then the reports then dropped. So, overall, public security also remains as low-level threat.

The graph below shows four security areas experience, sector wise and from 2017 to 2020 in a comparative picture:

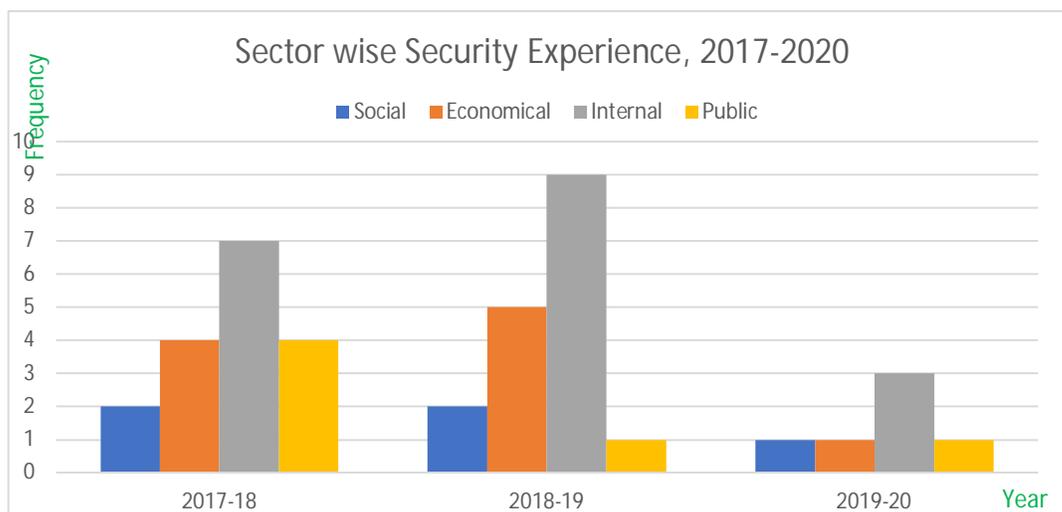

**Graph**: sector and year wise comparison of four security areas experience

The graph above shows that immediately after the Rohingya influx in 2017, there were a rising trend of crime rates in the four categories. In 2018, social and public security reports dropped while economic and internal security reports rose a bit. However, in 2020, reports in all four areas dropped. The law enforcement agencies did a great job.

The second graph (below) shows the trend of the security experience reports over three years, 2017 to 2020.

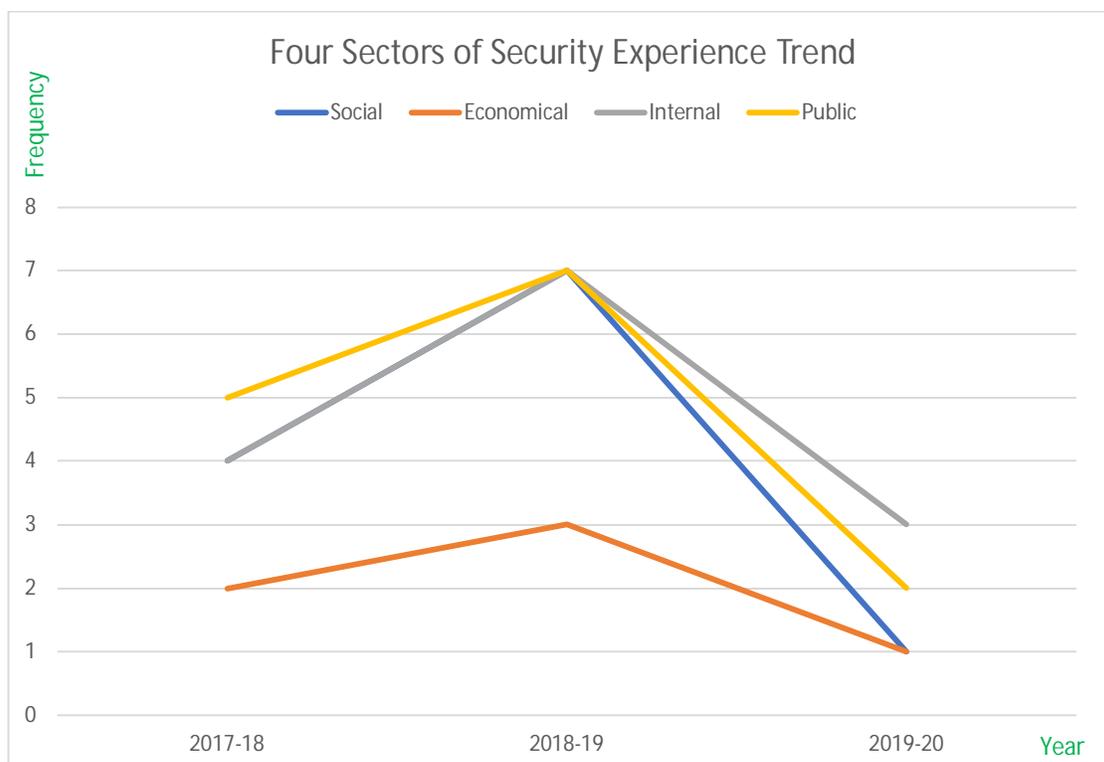

**Graph**: Fluctuations of news coverings of four security areas from August 2017 to August 2020.

The graph above shows that all the four security threats experience started in 2017 and they got worse in 2018, however, they steadily decline afterwards. The social, internal, and public security all got worse from August 2018 to August 2019 however, all of them improved in the following year. The law and enforcement agency played a great role in maintaining and controlling the situation.

The Rohingyas have been grateful to Bangladesh and its government for giving them shelter when their own state was killing them.[5] The sentiment of gratitude they are not willing to get involved into any activities that would harm Bangladesh somehow. A similar point was made by a group inside a refugee camp following a clash that they don't think of attacking the civilians because they are thankful for Bangladesh's generosity in sheltering the refugees (Reuters 2018). The law enforcement agencies of Bangladesh did a great job which reflects the gradual reduction of the rate or frequency of news reporting the incidents.

Another interesting finding is that the reports depict the threat to be serious for the country in the future whereas no present incident hints that that is currently any serious concern or threat. There have been 10 such reports in the 3 years. Gowher Rizvi, foreign policy advisor to the prime Minister, said that the Rohingya crisis is not only a threat to Bangladesh but also to the entire region, speakers at a seminar have (Abdullah 2019). The Prime Minister, Sheikh Hasina herself said that the Rohingyas become national and regional threat (Tribune Desk 2019b). Similarly, Awami League Presidium Member Begum Matia Chowdhury

---

said huge Rohingyas have now turned into a threat for Bangladesh's economy, environment, and security (Abdullah 2019). Foreign Minister Dr AK Momen made the remarks that the Rohingyas have become a threat to the security, socioeconomic conditions and environment of the country (Bhuiyan 2020). MP Israfil Alam, the minister said, 20 to 25% of the total population in Cox's Bazaar is comprised of Muslims from the Rakhine province. This huge population could become a threat to national security in future (Tribune Desk 2017). Also, the security experts expressed deep concern for the issue (The Daily Star 2017). In this connection, I think Copenhagen School's security construction in a good framework to understand this, 'Speech Act' to play role in constructing security, in this case, the speech and interpretation of the experts as well as the newspapers/reports' narrative seem to play similar role.

Also, it needs attention that due to media coverage and concerted efforts made by a section of the civil society, the Rohingya refugees issue seem more related to IS terrorists and more to do with security aspects, however, we need to keep in mind that Rohingya issue an ethnic issue and act accordingly (The Daily Star 2017).

In sum, I do not see any serious threat or security implication in Bangladesh so far. I did not find security concern or threat issue to be frequently or repeatedly reported or explained that they are serious threat experience for Bangladesh. The has not been frequent news covering of the same type of (in) security events which signify that the occurrence are of low-level threat. The newspaper covering is mainly reporting what happened but not depicting as serious security threat. However, in many cases the security concerns are descried serious threat for the future, but the evidence shows a different picture. Therefore, this research findings show that The Rohingya - Bangladesh case is an exception (till now) while the existing scholarship shows a link with refugees overstay and host state (in) security. So, the study makes a case that refugees overstay, and resultant host states insecurity are not always 'conclusive' or cannot be taken for granted. The findings ultimately add to questioning such over generalization put forth by the existing literature. Therefore, my research emphasizes on the need for a case-by-case analysis and evaluation.

## VII. Limitations of the study

I cover only two national dailies for three years period. I plan to cover more newspaper in near future making the project more rigorous. However, Though I covered only two national dailies, I think they cover all the important conflict incidents occurred during this time period.

Police file case data remains incomplete. Following a shooting of an army major in pretext of vehicle inspection in a police check post in Cox's bazar, the whole police department is undergoing a replacement procedure which hindered my data gathering. I plan on gathering all the police file cases over the period of three years in the future. Also, possibly such case files from another geographical/administrative area of Bangladesh to show a comparative analysis about crime rate and type with the area where Rohingyas are living. Also, I tried to create the threat level analysis table; I realize that the table may seem needing more work.

I excluded political impact of the Rohingyas, but the evidence suggests that they have political impact as well. Speakers in a roundtable revealed that in the Cox Bazar area Rohingyas are involved political activities. Many local politicians, for instance, local members and chairmen want to win elections with the votes of Rohingyas whereas only the citizens of Bangladesh could vote. Such scenario also explains that local politician also do not allow many Rohingyas to become registered refugees (The Daily Star 2017). This is also a kind of threat to the country. I also exclude natural resource or environmental security factors which are reported to have adverse impact due to Rohingyas refuge in Bangladesh.

Lastly, this is a study on three years, however, as long as large scale Rohingyas are in Bangladesh, the issue of host state security will persist, and it is possible that the situation may get worse as time passes, so, the security scenario or impact analysis requires constant and recurring investigating.

## VIII. Conclusion

Though I consider the four security areas separately, they can be understood as part of national security of Bangladesh. For instance, economic security is considered as national security (Navarro 2018a) and threat to economic security is a threat to national security (Navarro 2018b). So, the four security areas are important areas to consider. While there have been reports of insecurity issues, the findings suggest the over the period of three years, form august 207 to august 2020, Bangladesh experience only low level (in) security experience from social, economic, internal, and public security experience mostly remains in low level. The causes include the role of the government, the role local administration and law enforcement, as well as the international community's funding for various projects regarding fulfilling the basic needs of the refugees in Bangladesh contributed to this. This also come along with the fact that the majority of the Rohingyas are content and pleased as the government of Bangladesh or Bangladesh as a country gave them place to stay when their own government was killing them.

However, it must be clear that because this has not been a threat to security in the last three years, does not make any prediction about future. one of the most ostensible areas that has seemingly have impact is 'service sector; which may have negative impact on the GDP since it has record of impacting the GDP positively otherwise (M. M. Alam, Sultan, and Afrin 2010). What awaits for Rohingyas crisis solution remains uncertain even the United Nations Organization itself remains divided on the issue (Minar 2018c). The situation gets more complicated as the big powers are playing roulette to materialize their interest instead of solving Rohingyas crisis (Taufiq 2019a). Also, since the well-being of an actor or person (e.g., Rohingya) shapes their migration decisions which is in turn subject to the socio-economic environment (may be seen as stochastic dynamics), an actor/individual seeks for the optimal strategy under imperfect information and uncertainty. So, they are unlikely to stay or return to a place that has more socio-economic uncertainties (Pramanik 2021; Polansky and Pramanik 2021). This complicates Rohingyas decision to return in individual level. On top of that the regime in power has not made any proven progress so far to take them back to Myanmar. The COVID-19 pandemic adds to the uncertainty. While the current economic shock due to COVID-19 has not yet pushed the economy in a precarious position, the long-term shock can be highly disruptive which may contribute to worsen the situation (Ahamed 2021). Research also shows that a rigid Current Accounts position along with its slow adjustment and its inflexibility against real shocks, lower degree of capital mobility and a slow progress of capital market integration with the rest of the world, which shows mor e adverse impact on the COVID shock (M. M. Alam, Khondker, and Molla 2013). A lot depends on how things play out in the future, the role of international community, the role of the law enforcement agencies, and the role of the Rohingyas. The continuation of sufficient fund flow remains a concern. The security scenario inside the Rohingya camps is not good. In October 2020 (beyond by analysis timeframe), there has been a number of clashes between two groups inside the camps over asserting dominance at the Lombasia camp in Kutupalong of Ukhiya, Cox's Bazar (Aziz 2020a). The clash left 8 people dead, 50 injured and over 150 houses were set ablaze. At least 12 people have been arrested over the shootings at Ukhiya's Kutupalong Rohingya camp that killed eight people till now (Aziz 2020b). in the past it was evident that there have been facilities in Myanmar produce and export Yaba, a deadly drug, to the neighboring counties (Minar 2018b). These are deadly because drug crime is considered to pose a major threat to the national security of Bangladesh. (Minar 2018a).

Another point to observe is that the 'speech act' from the top governmental positions seem always cautionary, depicting the dangerous and emerging threat. Foreign minister, the foreign policy adviser to the Prime Minister, and the Prime Minister herself are always bold about the growing and gradual complicated nature of the Rohingyas refuge in Bangladesh and the possible security threat for Bangladesh and the region (Bhuiyan 2019).

So, the logic of prevention, prevention before security threats become security concerns dominate state of insecurity perception in Bangladesh. While this is an excellent approach, the evidence shows that till now Bangladesh has not experienced any serious security threat in the last three years. There are some criminal activities and offenses, but these are only low-level security threat at best. Therefore, this research presents empirical evidence that challenges conventional wisdom that the refugees are security threats or challenges to the host states.

**Appendix A: Indicators of Four Security Threat Areas:**

| Security Threat Areas | Threat Indicators |
|---|---|
| **I. Social Security** | • Threat to socio-cultural unity of host state, Bangladesh<br>• Rohingyas mixing with local people<br>• Rohingyas spreading throughout the country<br>• Rohingyas getting legal documents in the country, for instance, National Identity Cards, Passports (through illegal means) |
| **II. Economic Security** | • Decrease wages for local people<br>• Taking away jobs of local or national people resulting unemployment and job insecurity of the Bangladeshi nationals<br>• Competition for local business of Bangladeshi nationals (unfair competition)<br>• Putting burden on the economy of Bangladesh and harming economic growth |
| **III. Internal Security** | <ul><li>Linkages to terrorism and terrorist activities</li><li>Violent criminal activities</li><li>Involvement with national and international crimes (e.g., drug business, small and light weapons business etc.)</li><li>Pose challenge to law-and-order situation</li></ul> |
| **IV. Public Security** | <ul><li>Threat to peoples' wellbeing, and social welfare</li><li>Rising criminality causing wellbeing of people</li><li>Threat to social and public order</li><li>Strain on locals' education, transport, and communication facilities</li></ul> |

**Appendix B: Security Threat Level Assessment Table:**

| Security Threat Areas | Threat Level Indicators |
|---|---|
| **I. Low Level Security Threat** | <ul><li>Affects only an individual or small groups of up to 15 people</li><li>Very limited adverse effect, only local implication</li><li>Threat type is usual same as other parts of the country is facing, for example, petty crimes, person to person tussle etc.</li><li>Easily solvable by local authorities</li></ul> |
| **II. Medium Level Security Threat** | <ul><li>Affects small group of 15-30 people (Group level incident)</li><li>Limited adverse effect, only local implication</li><li>Threat type is usual same as other parts of the country is facing, for example, petty crimes, person to person tussle etc. (the scope may be bigger)</li><li>Solvable by local authorities and law enforcement agencies</li></ul> |
| **III. High Level Security Threat** | <ul><li>Affects large groups of more than 30 people</li><li>Adverse effect goes beyond local context and have national level implication</li><li>Unprecedented/unique type of threats with possible national level implication law and order seems vulnerable</li><li>Challenging for the law enforcement agencies to solve it</li></ul> |